# Analogue two-dimensional semiconductor electronics

Dmitry K. Polyushkin[1,†], Stefan Wachter[1,†,*], Lukas Mennel[1], Maksym Paliy[2], Giuseppe Iannaccone[2], Gianluca Fiori[2], Daniel Neumaier[3], Barbara Canto[3] and Thomas Mueller[1,*]

[1] *Vienna University of Technology, Institute of Photonics, Gußhausstraße 27-29, 1040 Vienna, Austria*
[2] *Dipartimento di Ingegneria dell'Informazione, Università di Pisa, Via Caruso 16, 56122 Pisa, Italy*
[3] *AMO GmbH, Otto-Blumenthal-Straße 25, 52074 Aachen, Germany*

† These authors contributed equally to this work.
*Corresponding authors: stefan.wachter@tuwien.ac.at, thomas.mueller@tuwien.ac.at

**While digital electronics has become entirely ubiquitous in today's world and appears in the limelight, analogue electronics is still playing a crucial role in many devices and applications. Current analogue circuits are mostly manufactured using silicon as active material, but the ever present demand for improved performance, new devices and flexible integration has – similar to their digital counterparts – pushed for research into alternative materials. In recent years two-dimensional materials have received considerable research interest, fitting their promising properties for future electronics. In this work we demonstrate an operational amplifier – a basic building block of analogue electronics – using a two-dimensional semiconductor, namely molybdenum disulfide, as active material. Our device is capable of stable operation with good performance, and we demonstrate its use in feedback circuits such as inverting amplifiers, integrators, log amplifiers, and transimpedance amplifiers.**

At a system level, electronic devices can be characterized as either analogue or digital. While digital electronics works by using strictly defined, discrete signal values – 0 and 1 – in analogue electronics a signal can take any physically available level. Although the incredible increase in performance/price ratio of digital circuits has made many kinds of analogue circuits obsolete, there is still significant demand for analogue electronics in today's world. As with their digital counterparts, manufacturing of analogue electronics is still mostly done on silicon, but also here the never ceasing demand for higher performance, new kinds of devices and different, flexible integration is pushing research into new materials[1–4]. Two-dimensional (2D) materials, which due to their inherent thinness exhibit highly interesting properties in the electrical, optical and mechanical domain, have garnered a tremendous amount of research interest[5]. The semiconducting transition metal dichalcogenides (TMDs) in particular show great promise for future electronics.[6–11] So far, however, most research regarding electronic devices using semiconducting 2D materials has been focused mainly on digital electronics, mostly limited to small circuits[12,13]. Often devices are implemented purely for demonstration purposes of an underlying transistor architecture[14] and in part limited by a lack of large scale synthesis. During the recent years significant process has been achieved in high-quality large-scale growth of 2D materials[9,15–21] and circuits no longer limited to small area exfoliated flakes are starting to emerge. [22–27]

The research invested in carbon nanotubes (CNTs) also focuses mainly on digital electronics[28,29] and demonstrations of analogue electronic devices are only slowly beginning to appear[30]. The main challenge this technology is still facing is the efficient sorting of CNTs of same chirality, which is required to be very close to perfect to achieve high on/off ratios and consistent transistor performance[31].

While graphene as well as $MoS_2$ have shown considerable success in the implementation of RF circuits[2,32], traditional basic building blocks of analogue electronics have so far been lacking. One such basic building block is the operational amplifier (OPA). This kind of device with two inputs and one output, amplifies the difference in potential at the inputs with a very high gain factor already at zero frequency. What makes OPAs a fundamental building blocks in analogue electronics is their versatility. Using feedback circuits, mainly composed of simple elements like resistors or capacitors, the properties of the overall circuit like gain, frequency response, etc. are mostly depending only on the external components. The properties of such circuits are only affected by variation in the performance of the OPA itself (for example due to tolerances and mismatch during manufacturing) to a minor degree. Using these feedback circuits, numerous functionalities can be implemented using the same OPA, ranging from simple amplifiers, to more complex systems like adders, integrators, differentiators, buffers, filters and many more.[33] From this point of view, OPAs are the crucial element toward the highly desirable and long-pursued goal of obtaining an all-2D materials based circuit, overcoming the actual limitation in many applications envisioned and already demonstrated by the 2D materials community, where, for example, sensors based on 2D materials are still interfaced with crucial components (e.g. for signal processing/treatment, buffering, etc.), externally supplied by traditional silicon-based devices.[34–38]

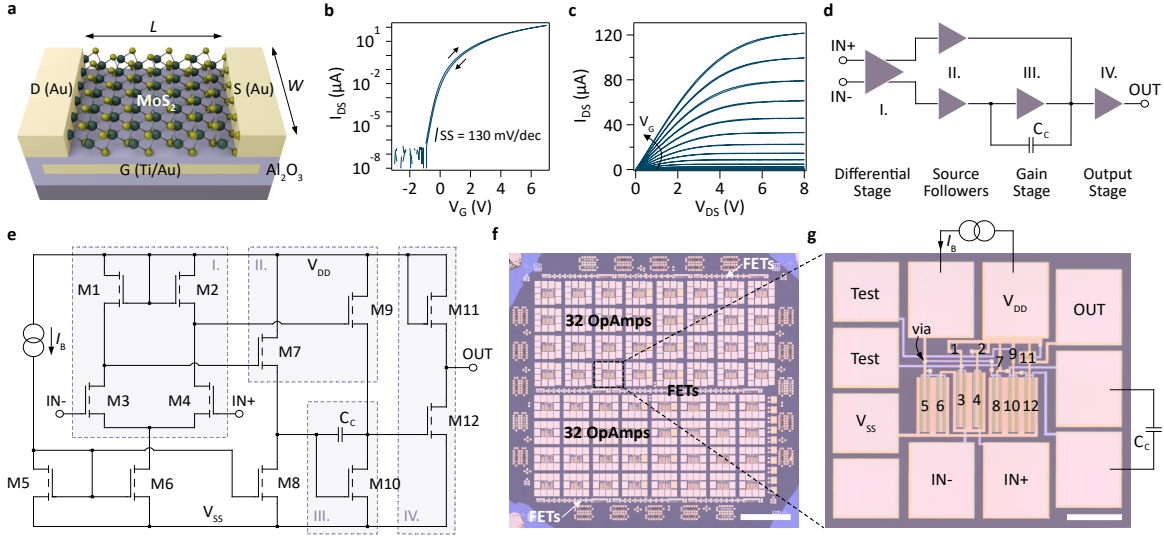

**Figure 1 | Operational amplifier circuit and fabricated chip. a**, Schematic representation of the back-gated transistor architecture. **b**, **c**, Transfer and output characteristics of a typical transistor on the chip (W/L = 4). The curves were acquired by scanning the voltage forth and back. **d**, **e**, Block- and transistor-level representation of the circuit, highlighting the different stages of the OPA: (I) differential input stage, (II) level shifters, (III) main gain stage with external capacitor used for phase-compensation, (IV) output stage. **f**, Optical micrograph of the chip, consisting of 64 OPAs and test FETs. Scale bar, 1 mm. **g**, Zoomed view of a single OPA showing the pinout and transistor labeling in accordance to (e). Scale bar, 100 μm.

## Operational amplifier design and fabrication

Our device consists of bottom-gated n-channel field-effect transistors (FETs) that use 2D semiconducting $MoS_2$ grown by chemical vapour deposition (see Methods and Supplementary Information, Figs. S1 and S2 for details about the growth and film itself)[39], as schematically shown in Fig. 1a. The silicon wafer with dry oxide serves only as a carrier substrate and can in principle be replaced by any non-conductive rigid or flexible substrate. All lithography steps are done by e-beam lithography. Gate, source and drain metal contacts are fabricated using an e-beam evaporation system. The gate dielectric, $Al_2O_3$, is deposited using atomic layer deposition (ALD), and definition of VIAs and $MoS_2$ channels is done via wet- and dry-etching. A detailed summary of the fabrication process can be found in the Methods section. Optical micrographs of the whole chip and a single OPA are shown in Figs. 1f and g, respectively. The chip contains a total of 64 OPAs, with a foot print of ~0.04 mm² each (without bonding pads), and additional test structures.

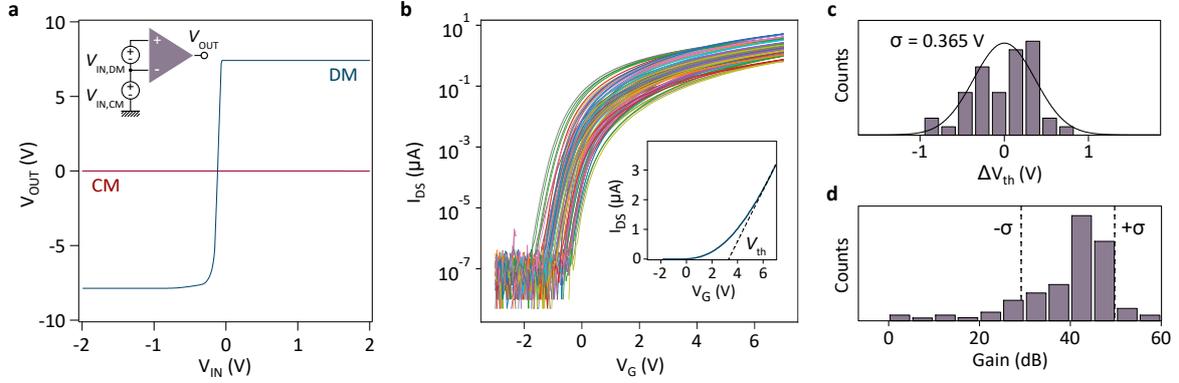

**Figure 2 | Device uniformity. a**, Cadence Spectre simulated large-signal sweep gain of the OPA (DM, differential-mode; CM, common-mode). **b**, Measured transfer characteristics of 46 transistors spread over the entire chip area ($V_{DS}$ = 100 mV). The inset shows a linear fit of the transfer characteristic to determine the threshold voltage. **c**, Histogram of the threshold voltage values extracted from the transfer characteristics of the data in (c). **d**, Distribution of low-frequency gain values determined by Monte-Carlo simulations of the entire OPA circuit, based on the device statistics in (b).

Measurements of single transistors show excellent performance with small hysteresis, on/off ratios exceeding 8 orders of magnitude, mobilities peaking around 10 cm²/Vs with values reaching up to almost 20 cm²/Vs (see Supplementary Information, Fig. S3), $SS$ = 130 mV/dec sub-threshold swing, and adequate saturation (Figs. 1b, c). All FETs are of enhancement-mode type with a threshold voltage of $V_{th} \approx 3.2$ V (see Figs. 1b and 2b). Since they are to be used in a complex electrical circuit, reproducibility of the main electrical parameters represent a fundamental issue. To this purpose, we characterized transistors with identical width-to-length ratios ($W/L = 4$) on the full area of our chip (~6×6 mm², sample size $N = 46$ devices) and extracted the threshold voltage and the charge carrier mobility. For the threshold voltage $V_{th}$ we find a standard deviation of $\sigma V_{th} = 0.365$ V (Fig. 2c). Confining the characterization to a smaller area ($N = 6$), more comparable to the actual size of our OPA circuit, yields a variation of $\sigma V_{th} = 0.13$ V. Scaling these values for the equivalent oxide thickness (EOT) of 0.5 nm, as targeted as by the IRDS (the successor to the now retired ITRS)[40], these values would correspond to roughly $\sigma V_{th} = 11$ mV for the large area variation of the threshold voltage – a value comparable to what can currently be achieved with silicon technology.[41,42] In contrast to silicon technology, which is fundamentally limited by thickness variations and dangling bonds when scaled down aggressively[8,40], this value is expected to improve with the ongoing developments of growth, transfer and general processing technologies for large scale 2D materials. Examples of such an improvement might be seen in improved dielectrics with EOT < 1 nm as recently demonstrated[43,44], more mature encapsulation techniques[45,46], or contact engineering for efficient current injection[47].

The OPA circuit (Fig. 1e) consists of 12 transistors (M1–M12) of different $W/L$-ratios, which are the main design parameters in our devices. The amplifier employs a three-stage design, with the low-frequency gain of each stage being given by the square-root of the relative $W/L$-ratios of the load- and input-transistors[48,49]. As indicated in Fig. 1d, a differential input stage (I) with a voltage gain of $A_\text{I} = \frac{1}{2}\sqrt{(W/L)_3/(W/L)_1}$ that, through two (unity-gain) source-followers acting as level shifters (II), drives the main gain stage (III) with a gain of $A_\text{III} = \sqrt{(W/L)_{10}/(W/L)_9}$. The output-stage (IV) provides medium gain according to $A_\text{IV} = \sqrt{(W/L)_{12}/(W/L)_{11}}$ while reducing the output resistance. The voltage gain of the complete amplifier is $A_\text{tot} = A_\text{I} A_\text{III} A_\text{IV}$. A bias current is supplied to the gain stages using Wilson current mirrors integrated into the circuit (transistors M5, M6 and M8). While typically supplied from an integrated and self-compensated current source, we supply this current externally in order to allow for a simpler circuit design as well as device-by-device adjustment of the bias for optimal performance. This current source could also be replaced by an external resistor. The amplifier was designed for $A_\text{I} = 13$ dB, $A_\text{III} = 22$ dB and $A_\text{IV} = 8$ dB, resulting in a total gain of $A_\text{tot} = 43$ dB. This yields the transistor parameters summarized in Tab. 1.

| Transistor | $W$ (µm) | $L$ (µm) | W/L |
|---|---|---|---|
| M1, M2 | 5 | 10 | 0.5 |
| M3–6, M8, M10, M12 | 200 | 5 | 40 |
| M7, M9 | 5 | 20 | 0.25 |
| M11 | 30 | 5 | 6 |

**Table 1.** Nominal widths and lengths of the transistors M1–M12 in the circuit.

In order to gauge the viability of the circuit we used Cadence Spectre to model our transistors according to the statistical data in Fig. 2b (see Supplementary Information, Fig. S4 and S5). Since a complete model of backgated 2D semiconductor FETs is still not readily available, we fitted the experimental results with an EKV model[50] (see Methods) both in the subthreshold ($V_G$ < 3.2 V) and the inversion ($V_G$ > 3.2 V) regimes. Since all transistors operate in the inversion regime, we used the inversion model to simulate the OPA, obtaining a nominal low-frequency gain value of $A_{tot}$ = 42.9 dB – very close to the designed value, while confirming the viability of the transistor model and the circuit design. Since these calculations assume perfect matching of the transistors, we performed Monte-Carlo simulations with 1500 traces, employing the strong inversion model, in order to estimate the robustness of our circuit regarding mismatch using a threshold variation of $\sigma V_{th}$ = 0.365 V. Additionally, a mobility variation of $\sigma \mu$ = 5.5 cm$^2$/Vs, as extracted by comparison of the experimental data with DC Monte-Carlo simulations (81 traces, $V_{DS}$ = 100 mV, random sequence generation; Supplementary Fig. S4), was introduced. This yields a mean value of the dB gain of 39.5 dB with a standard deviation of 10.3 dB, which is reasonably close to the nominal values (Fig. 2d). This, however, again highlights the importance of uniformity of the underlying elements in complex circuits, as well as a more fault tolerant design usually employed in industrial manufacturing.

**Operational amplifier performance**

Electrical measurements to characterize the fabricated devices were performed in a vacuum probestation (~10$^{-6}$ mbar) to avoid influence of atmospheric adsorbates on transistor performance. The bias current was set using an external current source and adjusted for best performance of the device around typical values of $I_B$ = 5–10 µA. Once adjusted, the bias current was kept constant and consistent operation of the device was stable for several hours without degradation or drifts in performance. Large-signal DC (f = 1Hz) sweeps were performed to determine the characteristics of the OPA in static condition. Using a supply voltage of $V_{DD}$/$V_{SS}$ = +/-10 V and keeping the negative input IN– grounded, the voltage at the positive input IN+ was swept twice from negative to positive voltage and back (differential-mode, as illustrated in Fig. 2a). Remarkably, the resulting curve does not exhibit any measurable hysteresis. The swing of the output voltage shows a gain of 33 dB (see blue line in Fig. 3a) and a large voltage swing of almost 15 V, with the best device showing a low-frequency gain of 36 dB (see grey line in Fig. 3a). While these gain values are in agreement with the design, they are also smaller than what is typically obtained in commercial OPAs (>60 dB). Achieving higher gain values using the circuit in Fig. 1e would require excessively large $W/L$-ratios for M3, M4, M10 and M12. A possible solution could be to replace all load transistors with n-type depletion-mode FETs, operated as current sources, or using CMOS technology instead by employing a different 2D semiconductor than $MoS_2$.

Since an OPAs main function is to amplify the potential difference between the positive and negative input, concurrently sweeping both inputs should result a constant output. In real circuits, however, small variations during manufacturing will cause asymmetries in the circuit that results in a non-zero slope of the output voltage. Such a common-mode (CM) measurement (illustrated in the inset in Fig. 2a) was performed showing only insignificant changes in output voltage (see red line in Fig. 3a) and we extract a common-mode rejection ratio of ~50 dB.

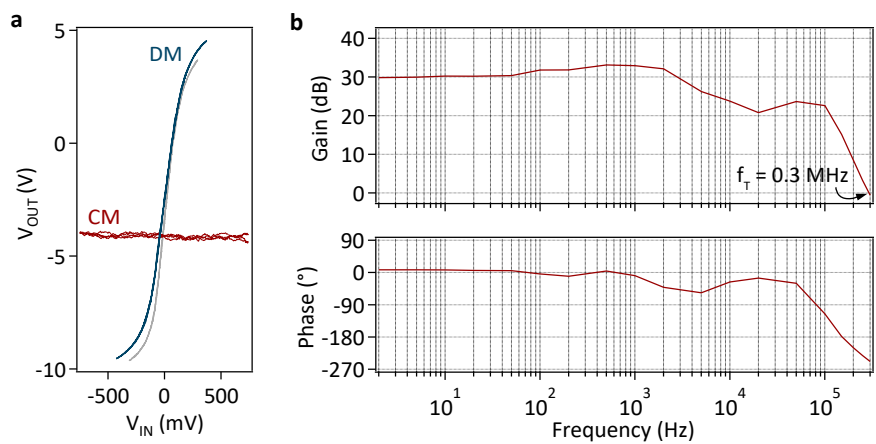

**Figure 3 | Operational amplifier performance. a**, Measured large-signal DC gain of two different OPA devices in differential-mode (DM) and common-mode (CM). The curves were obtained by sweeping the input voltage twice forth and back. **b**, Bode-plot of the small-signal gain and phase of the OPA showing a maximum gain of 33 dB and a unity-gain frequency of 0.3 MHz.

The frequency response of the device was characterized using low-amplitude (100 mV peak-to-peak) sine waves of varying frequency as input in differential-mode configuration and plotting gain and phase shift of the output with regard to the input (see Fig. 3b). These measurements yield a maximum gain of 33 dB, rolling off at around 5 kHz, and a unity-gain transition frequency of $f_T = 0.3$ MHz. To avoid unwanted oscillation of the OPA in typical negative feedback circuits (as demonstrated later) the phase margin (i.e. the distance of the phase to 180° at 0 dB gain) should practically be 60° or larger. According to Fig. 2b this requirement is clearly not fulfilled, because each of the three stages in our circuit contributes 90° to the total phase shift. This, however, can be compensated by a technique called pole-splitting, which decreases the phase at the cost of a slight reduction in gain.[51] In our device this is accomplished by employing an external capacitor $C_C$. Details and results of this compensation can be found in the Supplementary Information, Fig. S6.

**Analogue electronic circuits**

Although the characterization is performed in open-loop circuits, in practical applications OPAs are used in closed-loop configurations, where the output is fed back into the positive or negative input. The functionality of the circuit will then mostly be determined by the components within the feedback network instead of the actual performance of the OPA. In this work we demonstrate several feedback circuits commonly employed in electronics – inverting amplifier, integrator, logarithmic amplifier, and transimpedance amplifier (see Figs. 4a-d). The feedback network is realized externally, using discrete components such as resistors, capacitors and diodes. Fig. 4a shows the implementation of an inverting amplifier with a nominal gain of $V_{OUT}/V_{IN} = -R_2/R_1 = -3.9$. Figs. 4b and c show the results of the integrator and logarithmic amplifier, that, as their names suggest, show the time integrated and the logarithm of the input voltage, respectively, at their outputs. A more applied example is shown in Fig. 4d where the OPA is used as a transimpedance amplifier that converts the current of an illuminated photodiode to an output voltage. The results nicely show the characteristic 100 Hz blinking of the classic incandescent bulb used as a light source in the measurement. These experiments demonstrate that our kind of device can be used in a plethora of applications already in this very prototypical stage. While still work in progress, with the expected improvements in processing of 2D materials, manufacturing of such devices on an industrial scale seems possible.

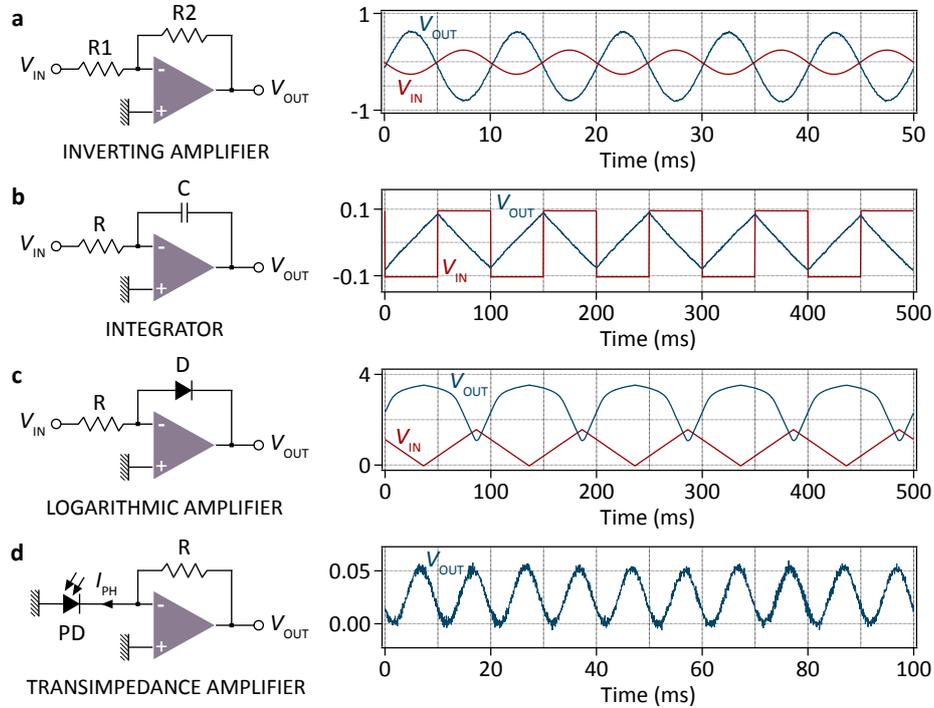

**Figure 4 | Analogue electronic circuits. a**, Inverting amplifier, **b**, integrating amplifier that time-integrates the input signal, **c**, log-amplifier that displays the logarithm of the input signal at the output, **d**, transimpedance-amplifier converting a current-input signal (in this experiment from a photodiode illuminated by an incandescent bulb, hence the characteristic 100 Hz signal) to a voltage output-signal.

**Conclusion**

In this work we demonstrated the design, fabrication and characterization of an OPA based on n-type enhancement-mode FETs using a 2D semiconductor (MoS$_2$) as active material. In addition, we show the characterization of the underlying n-channel transistors of remarkably low variability, comparable to that achieved by silicon technology. This low variability marks a crucial point in our work, since any design and implementation of a larger, more complex circuit, such as an OPA, places stringent requirements on the uniformity of the constituent components. While the gain of the OPA is not as high as in commercial silicon devices, we show the applicability of our OPA in typical feedback circuits. We believe that a more fault tolerant design, improvements in processing and material quality, and, most importantly, the development of CMOS technology in 2D materials might be a viable path towards application of 2D semiconductors in analogue electronics.

## METHODS

### CVD growth of MoS$_2$

MoS$_2$ was grown on c-plane sapphire substrates using a two-zone tube furnace. Powdered sulfur precursor (Sigma Aldrich) was evaporated at atmospheric pressure under constant argon flow at ~150°C. Downstream from the sulfur source the sapphire substrate was placed above the MoO$_3$ powder (Sigma Aldrich) precursor which was evaporated at ~700°C. After a growth time of 10 minutes the furnace was left to cool down naturally and the finished substrate with grown film extracted.

### Circuit design and modeling

Early stage circuit modeling and design was done using LTSpice employing a model fit to the data from single transistor measurements. Since no circuit model is available for transistors based on 2D materials, we resorted to the EKV model, which is solidly physically based and requires few parameters, but is tailored to silicon field-effect transistors. For this reason we implemented and tuned in Cadence Spectre two different EKV model: one (the "inversion" model) that is very accurate in reproducing the experimental transistors in inversion and loses some accuracy in subthreshold conditions; a second model (the "subthreshold" model) that is very accurate in reproducing the experimental transistor characteristics in subthreshold conditions and loses some accuracy in inversion. The main difference between two models lays in the higher channel doping concentration value of $N_\text{D} = 2 \times 10^{18}$ cm$^{-3}$ for the subthreshold model ($V_\text{G} < 3.2$ V), the related value for the inversion model ($V_\text{G} > 3.2$ V) was $0.1 \times 10^{18}$ cm$^{-3}$. In the specific case of the OPA simulation all transistors are in inversion, so that the inversion model has been used. A statistic block was added to the model by considering both process and mismatch variations due to the threshold voltage with a sigma $\sigma V_\text{th} = 0.365$ V, value given from measurements, and mobility $\sigma\mu = 5.5$ cm$^2$/Vs, derived from DC Monte-Carlo simulations, instead. The OPA's characteristics, as well statistics were then extracted by simulations in Cadence Virtuoso.

### Device fabrication

All lithography steps are done by e-beam lithography using a Raith e-line system and Allresist PMMA 679.004. Metal deposition is done in a Leybold e-beam evaporation system at a pressure < $3\times10^{-7}$ mbar. The first metal layer defines the gate electrodes (and parts of the routing) and consists of 3/25nm of Ti/Au. Afterwards the gate dielectric (30 nm, Al$_2$O$_3$) is deposited using atomic layer deposition from TMA and water at 200°C in which we subsequently use 30% KOH solution to wet-etch VIAs in order to facilitate contact of bottom and top-metal where desired. The $MoS_2$ film is lifted off the growth substrate by spinning of thick polystyrene (PS), and a short immersion in KOH which lifts off the polymer/$MoS_2$ stack, which is then rinsed in DI-water. After drying at slightly elevated temperatures the film is transferred in a dry-air glovebox on the prebaked target substrate which is heated slowly from room temperature to ~150°C in order to soften the PS and facilitate proper contact of the film to the substrate. The PS is subsequently dissolved in toluene. We use Ar/SF6 plasma etching in an Oxford Cobra RIE system to define the actual transistor channels. In a final step we deposit 30 nm Au which defines the source and drain electrodes and finalizes the routing of the circuit.

**Electrical measurements**

All measurements were performed in a Lakeshore TTPX probestation at room temperature and in vacuum (~$10^{-6}$ mbar). Single transistor characterizations were performed using an HP 4155C semiconductor parameter analyzer. Complete device measurements used Keithley 2614B SMUs as power supply and biasing sources. Input and output voltages were supplied and measured by an Agilent 33220A function generator and a Keysight Infiniivision oscilloscope, respectively. Due to the high capacitive load imposed on the circuit by the measurement setup, the contacting-needle used to measure the output voltage was actively buffered to enable measurements at high frequencies.

**ACKNOWLEDGMENTS**

We acknowledge financial support by the European Union (grant agreements No. 785219 Graphene Flagship and No. 796388 ECOMAT ), the Austrian Science Fund FWF (START Y 539-N16), and the Italian MIUR (FIVE 2D).

**AUTHOR CONTRIBUTIONS**

T.M. conceived the project. S.W. and T.M. designed the chip. D.K.P. grew the $MoS_2$ film. D.K.P. and S.W. fabricated the samples and performed the measurements. D.N. and B.C. contributed to the sample fabrication. L.M. characterized the $MoS_2$ film. G.F., M.P., and G.I. performed the Monte-Carlo simulations. S.W. and T.M. wrote the manuscript. All authors discussed the results and contributed to the manuscript.

**COMPETING FINANCIAL INTERESTS**

The authors declare no competing financial interests.